\documentstyle[amssymb,preprint,aps]{revtex}

\begin{document}
\draft
\title{Nonequilibrium transport through a quantum dot weakly coupled to Luttinger
liquids}
\author{{\ Yi-feng Yang and Tsung-han Lin}$^{\ast }$}
\address{{\it State Key Laboratory for Mesoscopic Physics and }\\
{\it Department of Physics, Peking University,}{\small \ }{\it Beijing}\\
100871, China}
\maketitle

\begin{abstract}
We study the nonequlibrium transport through a quantum dot weakly coupled to
Luttinger liquids (LL). A general current expression is derived by using
nonequilibrium Green function method. Then a special case of the dot with
only a single energy level is discussed. As a function of the dot's energy
level, we find that the current as well as differential conductance is
strongly renormalized by the interaction in the LL leads. In comparison with
the system with Fermi liquid (FL) leads, the current is suppressed,
consistent with the suppression of the electron tunneling density of states
of the LL; and the outset of the resonant tunneling is shifted to higher
bias voltages.\ Besides, the linear conductance obtained by Furusaki using
master equation can be reproduced from our result.
\end{abstract}


PACS numbers:73.63.-b,71.10.Pm, 73.40.Gk, 73.23.-b,

\baselineskip 20pt 
\newpage

Tunneling phenomena in mesoscopic systems have been intensively studied for
years both experimentally and theoretically. Since early 1990s, the
tremendous progress in nanofabrication and the discovery of novel
one-dimensional (1D) materials, especially the carbon nanotubes,\cite{iiji}
have led to a revived interest in studying the tunneling in systems
including 1D conductors.\cite{egger}\cite{morp}

It is well-known that at low temperatures the two-dimensional (2D) and
three-dimensional (3D) interacting electron systems can be described by
Landau's Fermi liquid (FL) theory. However, the interacting electron systems
in one dimension (1D) is quite different, it behaves as Luttinger liquid
(LL),\cite{tomo}\cite{lutt}\cite{hald} for that theory predicts a series of
unique characters, e.g., the absence of Landau quasiparticles, spin-charge
separation, suppression of the electron tunneling density of states (DOS),
anomalous power laws for transport coefficients with interaction-dependent
exponents, etc.\cite{voit}\cite{rao} Naturally, one may ask whether the
(resonant) tunneling between LLs is different from that of FLs.

Recently, Furusaki reported a theoretical study of the resonant tunneling
through a quantum dot (QD) weakly coupled to two LL leads (hereafter as
LL-QD-LL).\cite{furu} Assuming that the tunneling is an incoherent
sequential tunneling, the author derived the linear conductance using the
master equation approach. Similar studies have been done by other authors
earlier.\cite{kane}\cite{furu2}\cite{cham} In this paper, we present an
investigation for a LL-QD-LL system. Differently, we consider the tunneling
as a coherent process and study the nonequilibrium current through the
system with a finite bias voltage applied on the two \ LL leads. By using
nonequilibrium Green function technique,\cite{haug} a general current
formula is derived. We find that both the current and differential
conductance are strongly renormalized by the interaction in the LL leads. In
current-voltage characteristics, the current is first suppressed, consistent
with the suppression of the electron tunneling density of states; and the
outset of the resonant tunneling is found to be shifted to higher bias
voltages, comparing to the system FL-QD-FL. Finally, the linear conductance
obtained by Furusaki using master equation can be reproduced from our result.

The Hamiltonian of the system can be split into three parts: $%
H=H_{leads}+H_d+H_T$, where $H_{leads}=H_L+H_R$ denotes the left and right
LL leads in its standard form (see \cite{furu}), $H_d=\sum_n\varepsilon
_nd_n^{+}d_n$ is the Hamiltonian of the QD, with $\{d_n^{+},d_n\}$ the
creation/annihilation operators of the n-th energy level in the dot (for
simplicity, the intradot Coulomb interaction has been neglected). $H_T$ is
the tunneling Hamiltonian and can be written as

\begin{equation}
H_T=\sum_{n,\lambda =L/R}(t_{n,\lambda }d_n^{+}\psi _\lambda +h.c),
\end{equation}
in which $\{\psi _\lambda ^{+},\psi _\lambda \}(\lambda =L/R)$ are the Fermi
operators at the end points of the left/right lead; $\{t_{n,\lambda }\}$ is
the tunneling constant. The spin indices have been suppressed since they are
not important in the following discussion.

Defining the mixed casual Green function $G_{n,\lambda }\left(
t_{1,}t_2\right) =-i\left\langle Td_n\left( t_1\right) \psi _\lambda
^{+}\left( t_2\right) \right\rangle _H$ and Green functions of all the other
types in the standard form,\cite{haug} the current from left LL lead into
the QD under a bias voltage $-V$ applied on the right lead can be evaluated
as follows (in units of $\hbar =1$):

\begin{equation}
J_L=e\left\langle \frac{dN_L}{dt}\right\rangle =-2e\sum_n%
Re \left[ t_{n,L}^{*}G_{n,L}^{<}\left( t,t\right) \right] ,
\end{equation}
where $G_{n,L}\left( \tau ,\tau ^{\prime }\right) =\sum_m\int_cd\tau
_1G_{n,m}\left( \tau ,\tau _1\right) t_{m,L}g_L\left( \tau _1,\tau ^{\prime
}\right) $ is the Dyson equation for contour Green function $G_{n,L}\left(
\tau ,\tau ^{\prime }\right) $, $G_{n,m}\left( \tau ,\tau ^{\prime }\right)
=-i\left\langle T_cd_n\left( \tau \right) d_m^{+}\left( \tau ^{\prime
}\right) \right\rangle _H$ is the full Green function of the QD, $g_L\left(
\tau ,\tau ^{\prime }\right) =-i\left\langle T_c\psi _L\left( \tau \right)
\psi _L^{+}\left( \tau ^{\prime }\right) \right\rangle _{H_L}$ is the local
Green function at the end point of the left LL lead without the coupling to
the QD, and $T_c$ is the contour-ordering operator.

After using Langreth theorem of analytic continuation, and the Fourier
transformation, the current can then be expressed as:

\begin{equation}
J_L=-2e\sum_{n,m}%
Re%
\int \frac{d\omega }{2\pi }\left[ t_{n,L}^{*}\left( G_{n,m}^r\left( \omega
\right) g_L^{<}\left( \omega \right) +G_{n,m}^{<}\left( \omega \right)
g_L^a\left( \omega \right) \right) t_{m,L}\right] .
\end{equation}

The Dyson equation for $G_{n,m}\left( \omega \right) $ is $G_{n,m}\left(
\omega \right) =\delta _{n,m}g_n\left( \omega \right) +\sum_lg_n\left(
\omega \right) \Sigma _{n,l}\left( \omega \right) G_{l,m}\left( \omega
\right) $, where $\Sigma _{n,l}\left( \omega \right) =\sum_{\lambda
=L/R}t_{n,\lambda }t_{l,\lambda }^{*}g_\lambda \left( \omega \right) $ is
the irreducible self-energy and $g_n\left( \omega \right) $ is the free
Green function of the dot without the coupling between the QD and leads.
Using Langreth theorem and noticing 
$\Sigma ^{r,a}=\pm \frac 12\left( \Sigma ^{>}-\Sigma ^{<}\right) $, the
current expression, Eq.(3), can be reduced to

\begin{equation}
J_L=e\sum_{n,m}\int \frac{d\omega }{2\pi }t_{n,L}^{*}\left\{ \left[ \left(
g^r\right) ^{-1}-\Sigma ^r\right] ^{-1}\left[ g_L^{>}\Sigma
^{<}-g_L^{<}\Sigma ^{>}\right] \left[ \left( g^a\right) ^{-1}-\Sigma
^a\right] ^{-1}\right\} _{n,m}t_{m,L}.
\end{equation}

Neglecting the off-diagonal tunneling matrix elements and their
energy-dependence, which is equivalent to take $t_{n,L}^{*}t_{m,L}=\delta
_{n.m}\left| t_{n,L}\right| ^2$,\cite{meir} then Eq.(3) reduces to

\begin{equation}
J_L=e\sum_n\left| t_{n,L}t_{n,R}\right| ^2\int \frac{d\omega }{2\pi }\frac{%
g_L^{>}g_R^{<}-g_L^{<}g_R^{>}}{\left( \left( g_n^r\right) ^{-1}-\Sigma
_{n,n}^r\right) \left( \left( g_n^a\right) ^{-1}-\Sigma _{n,n}^a\right) },
\end{equation}
which describes the nonequilibrium current in LL-QD-LL, and is the the
central result of this work. Similarly, one can easily obtain the current
from right LL lead to the QD, $J_R$, just by changing the subscript in
Eq.(5) $L\rightarrow R$, and easily to see $J_L+J_R=0$, the current
conservation, as expected.

As an important example, let us consider the quantum dot with only a single
energy level, i.e., $H_d=\varepsilon d^{+}d$. After taking a contour
integral, Eq.(5) reduces to

\begin{equation}
J=J_L=-J_R=e\left| t_Lt_R\right| ^2\frac{g_L^{<}\left( \varepsilon \right)
g_R^{>}\left( \varepsilon \right) -g_L^{>}\left( \varepsilon \right)
g_R^{<}\left( \varepsilon \right) }{\sum_{\lambda =L/R}\left| t_\lambda
\right| ^2\left| g_\lambda ^{<}\left( \varepsilon \right) -g_\lambda
^{>}\left( \varepsilon \right) \right| }+O\left( \left| t_Lt_R\right|
^2\right) .
\end{equation}
In which $g_L^{<,>}\left( \varepsilon \right) $ and $g_R^{<,>}\left(
\varepsilon \right) $ are the lesser (greater) Green functions of the left
and right leads without the couplings to the dot, and can be directly quoted
from:\cite{furu}

\begin{eqnarray}
g_L^{<,>}\left( \varepsilon \right) &=&\pm i\frac T{\left| t_L\right| ^2}%
e^{\mp \frac \varepsilon {2T}}\gamma _L\left( \varepsilon \right) , \\
g_R^{<,>}\left( \varepsilon \right) &=&\pm i\frac T{\left| t_R\right| ^2}%
e^{\mp \frac{\varepsilon -eV}{2T}}\gamma _R\left( \varepsilon -eV\right) , 
\nonumber
\end{eqnarray}
where $T$ is the temperature, and $V$ the bias voltage. $\gamma _{L/R}\left(
\varepsilon \right) $ is defined as $\gamma _{L/R}\left( \varepsilon \right)
=\frac{\Gamma _{L/R}}{2\pi T}\left( \frac{\pi T}\Lambda \right)
^{1/g_{L/R}-1}\frac{\left| \Gamma (1/2g_{L/R}+i\varepsilon /2\pi T)\right| ^2%
}{\Gamma \left( 1/g_{L/R}\right) }$, here $\Gamma \left( x\right) $ is the
Gamma function, $g_{L/R\text{ }}$ are interaction parameters characterizing
the left/right LL liquids, which should not be confused with the Green
functions with the same symbols, $\Gamma _{L/R}$ describes the effective
level broadening of the dot, proportional to $\left| t_{L/R}\right| ^2$ ,
and $\Lambda $ is the high-energy cutoff or a band width.\cite{furu}

In the lowest order approximation, Eq.(6) reduces to

\begin{equation}
J=eT\frac{\gamma _L\left( \varepsilon \right) \gamma _R\left( \varepsilon
-eV\right) }{\gamma _L\left( \varepsilon \right) \cosh \left( \frac %
\varepsilon {2T}\right) +\gamma _R\left( \varepsilon -eV\right) \cosh \left( 
\frac{\varepsilon -eV}{2T}\right) }\sinh \left( \frac{eV}{2T}\right) ,
\end{equation}
\ 

Eq.(8) describes the nonequilibrium current of LL-QD-LL under a finite bias
voltage. In the following, we present the numerical studies based on Eq.(8)
for the symmetric case, i.e., $g_L=g_R=g$, $t_L=t_R$ and $\Gamma _L=\Gamma
_R $.

Fig.1 shows\ the current vs bias voltage at a fixed value $\frac \varepsilon
{2T}=5$ for different interaction parameters of the LL leads, $g=0.9,0.7,%
\frac 12,\frac 13,\frac 1{10}$. The system with FL leads is also shown for
comparison (see the curve with $g=1$ in Fig.1). We find the following
features: (1) For FL-QD-FL , a steep increase occurs around $\frac{eV}{2T}=5$%
, which is consistent with the suppression of the electron tunneling of
states of the LL. The suppression becomes stronger with the decrease of $g$
(or the increase of the interactions in the leads). (2) The outset of the
resonant tunneling varies with the change of $g$, from $eV=\varepsilon $ for 
$g=1$ (i.e., the FL leads) to $eV=2\varepsilon $ for $g\thicksim 0$ (i.e.,
the strongest interacting LL leads). A current plateau appears again for $%
g\rightarrow 0$ around $eV=2\varepsilon $. (3) Somehow, the curves with $%
g\lesssim \frac 12$ intersect approximately at a point ($\frac{eV}{2T}=10$, $%
\frac J{J_{\max }}=\frac 12$), which may indicate the peculiar behavior for
strong interacting case.

The second feature above mentioned can also be seen from Fig.2, in which the
differential conductance vs $\frac \varepsilon {2T}$ for different
interaction parameters of LL, $g=1,\frac 12,\frac 1{10}$, is presented. The
location of the peak of conductance is shifted from $\varepsilon =eV$ to $%
\varepsilon =\frac{eV}2$ with the increase of $g$.

The nonequilibrium current as a function of $\frac \varepsilon {2T}$ at a
fixed bias is shown in Fig.3. Comparing to the current of FL-QD-FL, these
curves are strongly renormalized by the interactions of the LL leads:
instead a mesa for FL-QD-FL, a peak occurs at $\varepsilon =\frac{eV}2$; and
the half-width of the peak becomes narrower with the increase of the
interactions of electrons in the LL leads. The dependence of the peak
current on $T$ and $\varepsilon $ can be obtained by estimating the current $%
J$ at $\varepsilon =\frac{eV}2$. Because the asymptotic limit of the Gamma
function is $\Gamma \left( x\right) $ $\thicksim x^{x-\frac 12}e^{-x}\sqrt{%
2\pi }$ $\left( \text{ for }\left| x\right| \rightarrow \infty ,\text{and }%
\left| \arg x\right| \leq \pi -\delta ,\delta >0\right) $,\cite{grad} one
finds that at low temperatures $\left( T\ll \frac \varepsilon 2\right) $ the
current behaves as $J$ $\thicksim $ $\left( \frac \varepsilon {2\Lambda }%
\right) ^{1/g-1}$, whereas at high temperatures $\left( T\gg \frac %
\varepsilon 2\right) ,$ $J$ $\thicksim $ $\frac \varepsilon {2\Lambda }%
\left( \frac T{2\Lambda }\right) ^{1/g-2}$. Similar scaling forms have been
derived by Chamon and Wen by considering the incoherent sequential tunneling.%
\cite{cham}

It should be pointed out that our calculation is valid only when the higher
order contributions ($\gtrsim $ $O\left( \left| t_Lt_R\right| ^2\right) $)
can be neglected. The pole of the integrand in Eq.(4) satisfies the equation 
$\left( g^r\left( \omega \right) \right) ^{-1}-\Sigma ^r\left( \omega
\right) =0$. Neglecting the shift of the energy level of the QD, the lowest
order solution is $\omega =\varepsilon +%
Im%
\Sigma ^r\left( \varepsilon \right) $. Thus, the higher order correction can
be neglected only if $\left| \frac{\mathop{\rm Im}\Sigma ^r\left(
\varepsilon +\mathop{\rm Im}\Sigma ^r\left( \varepsilon \right) \right) -%
\mathop{\rm Im}\Sigma ^r\left( \varepsilon \right) }{\mathop{\rm Im}\Sigma
^r\left( \varepsilon \right) }\right| \ll 1$, or equivalently, $\left| \frac{%
\gamma \left( \varepsilon +\mathop{\rm Im}\Sigma ^r\left( \varepsilon
\right) \right) -\gamma \left( \varepsilon \right) }{\gamma \left(
\varepsilon \right) }\right| \ll 1$. Using the asymptotic limit of the Gamma
function, one finds that at low temperatures $\left( T\ll \frac \varepsilon 2%
,\frac{\left| eV-\varepsilon \right| }2\right) ,$ the condition reduces to $%
\frac \Gamma T\left( \frac \varepsilon {2\Lambda }\right) ^{1/g-1}\ln \left( 
\frac \varepsilon {2T}\right) \ll 1$ and $\frac \Gamma T\left( \frac{\left|
eV-\varepsilon \right| }{2\Lambda }\right) ^{1/g-1}\ln \left( \frac{\left|
eV-\varepsilon \right| }{2T}\right) \ll 1$; whereas at high temperatures $%
\left( T\gg \frac \varepsilon 2,\frac{\left| eV-\varepsilon \right| }2%
\right) ,$ $\frac \Gamma \Lambda \left( \frac T\Lambda \right) ^{1/g-2}\ll 1$%
. Therefore, our evaluations validate only when $T\neq 0$ and $\varepsilon
,eV,T\ll \Lambda $.\cite{expl}

Finally, as a quick check, we can obtain the linear conductance $\sigma $
from Eq.(7) by simply taking $V\rightarrow 0^{+}$, which gives:

\begin{equation}
\sigma =e^2\frac 1{2\cosh \left[ \frac \varepsilon {2T}\right] }\frac{\gamma
_L\left( \varepsilon \right) \gamma _R\left( \varepsilon \right) }{\gamma
_L\left( \varepsilon \right) +\gamma _R\left( \varepsilon \right) }.
\end{equation}
This is exactly the linear conductance obtained by Furusaki,\cite{furu} and
was confirmed by experiment recently.$\cite{ausla}$

In conclusion, we derive a general formula for nonequilibrium dc current
through a quantum dot weakly coupled to two Luttinger liquids. We consider
the tunneling as a coherent process and use nonequilibrium Green function
technique, different from Chamon and Wen and Furusaki. In symmetric case and
for the dot with a single energy level, we find that both the current and
the differential conductance are strongly renormalized by electron-electron
interactions in the LL leads. We hope that further experiments may provide
evidence for these theoretical predictions.

The authors would like to thank Guang-Shan Tian for many helpful
discussions. This work was supported by NSFC under grant No.10074001. One of
the authors (T.H.Lin) would also like to thank the support from the Visiting
Scholar Foundation of the State Key Laboratory for Mesoscopic Physics in
Peking University.

\smallskip $^{\ast }$ To whom correspondence should be addressed.

$
\bigskip $

\section*{Figure Captions}

\begin{itemize}
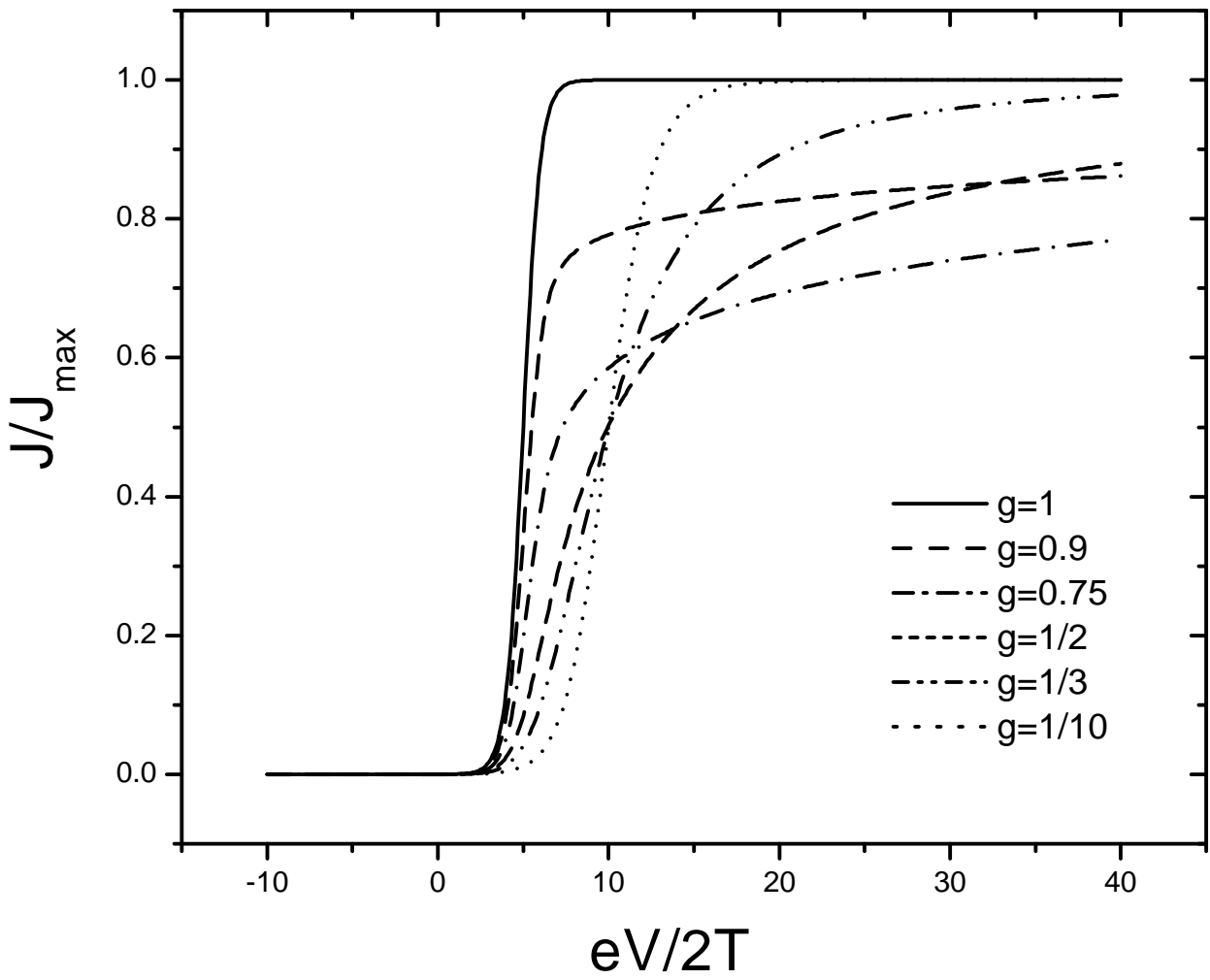
\item[{\bf Fig. 1}]  Current, scaled by its maximum at $\frac \varepsilon {2T%
}=5$, vs bias voltage for $g=0.9,0.75,\frac 12,\frac 13,\frac 1{10}$. $%
\varepsilon $ is the energy level of the dot, $g$ is the interaction
parameters of the leads. Curves for $g=\frac 12,\frac 13,\frac 1{10}$
intersect approximately at a point $\left( \frac{eV}{2T}=10,\frac J{J_{\max }%
}=\frac 12\right) $. The case with FL leads $\left( g=1\right) $ is also
shown for comparison..

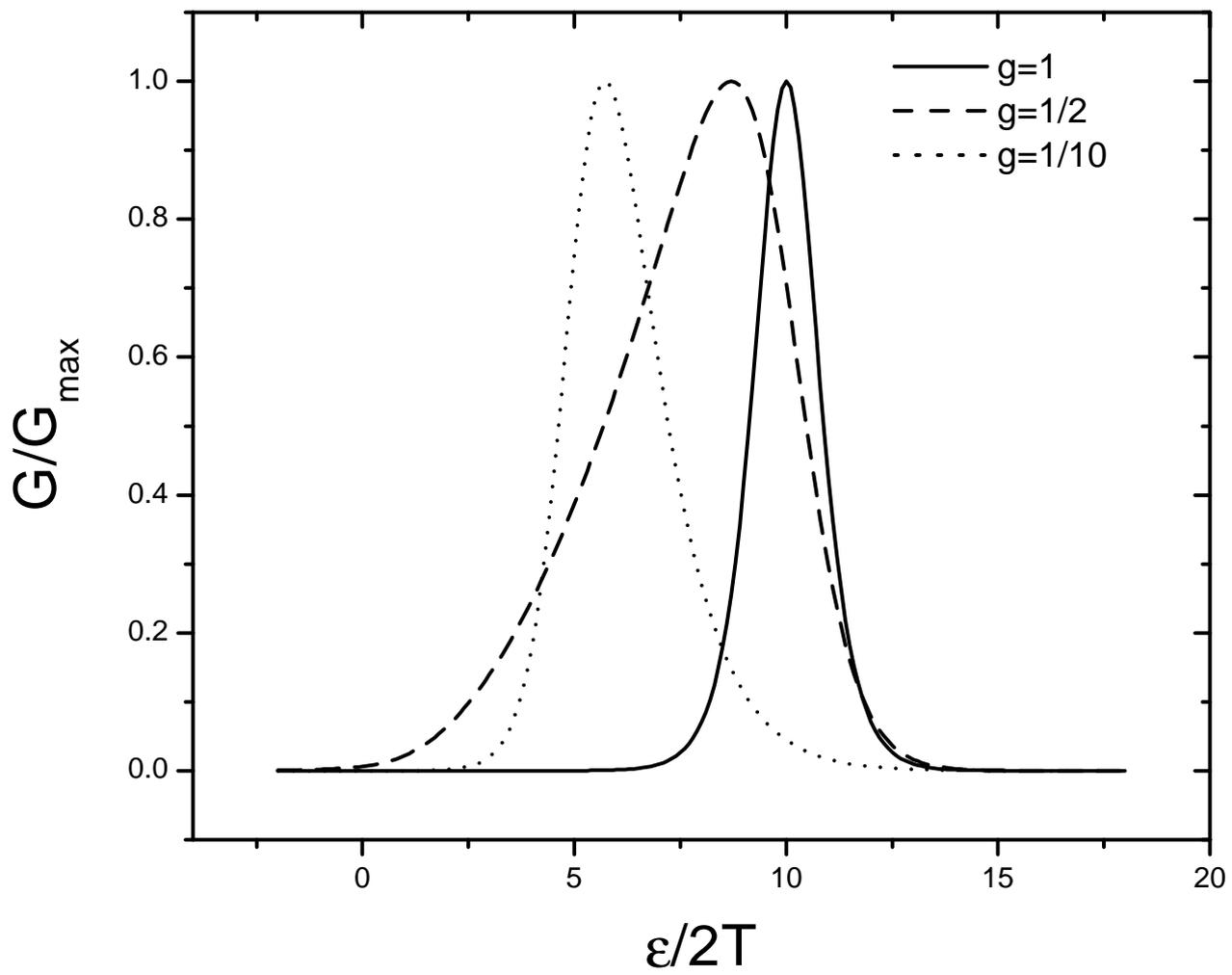
\item[{\bf Fig. 2}]  Differential conductance scaled by its maximum at $%
\frac{eV}{2T}=10$ as a function of $\frac \varepsilon {2T}$ for $g=1,\frac 12%
,\frac 1{10}$. The location of the maximum is shifted away from $\varepsilon
=eV$ for $g=1$ to $\varepsilon =\frac{eV}2$ for $g\sim 0$.

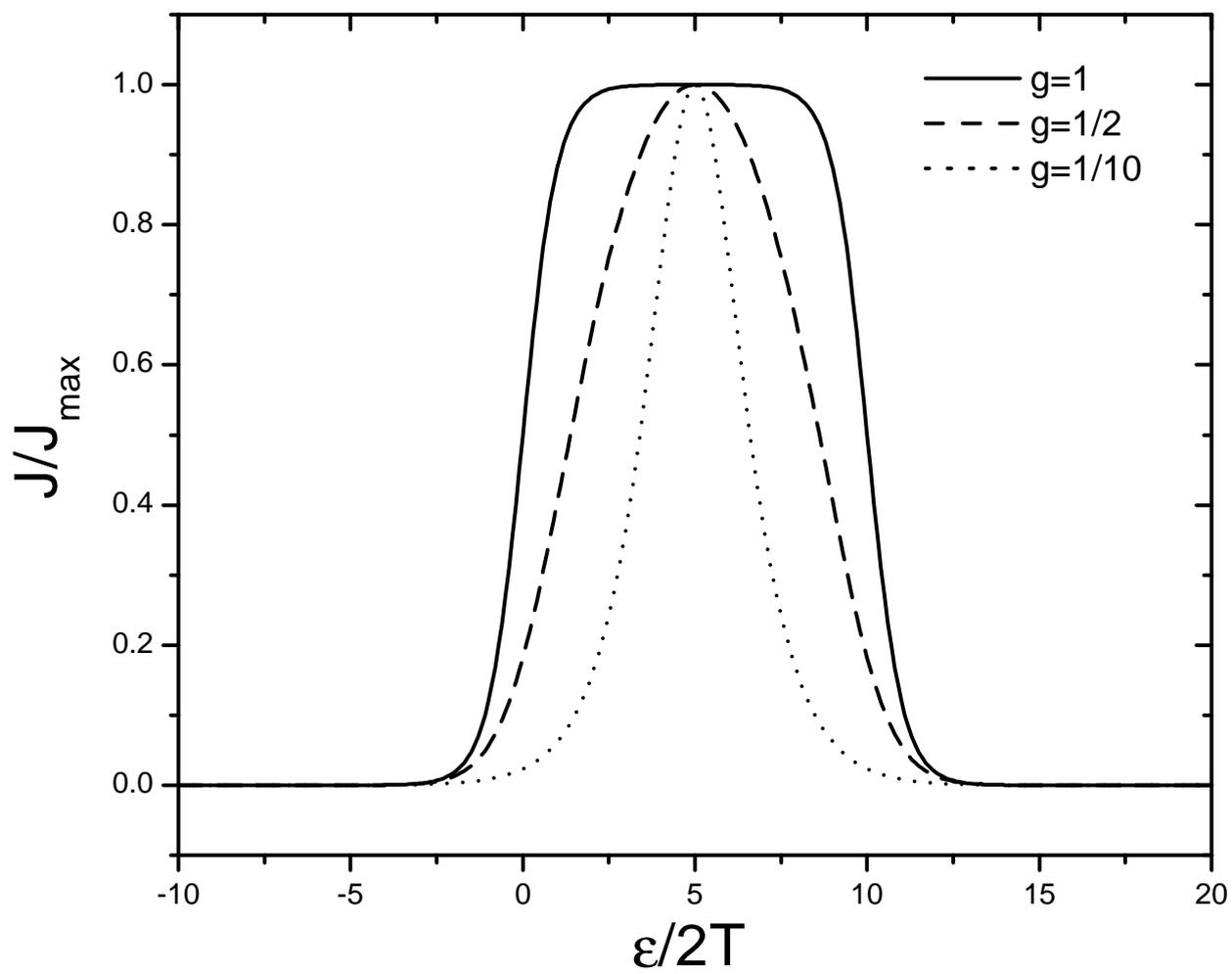
\item[{\bf Fig. 3}]  Current vs $\frac \varepsilon {2T}$ for $g=1,\frac 12,%
\frac 1{10}.$ The current is scaled by its maximum at bias voltage $\frac{eV%
}{2T}=10,$ For $g<1$(Luttinger liquid), a peak occurs at $\varepsilon =\frac{%
eV}2$, and the half width of the peak becomes narrower with the decrease of $%
g$. The curve with $g=1$ (Fermi liquid) is also shown for comparison.
\end{itemize}


\begin{references}
\bibitem{iiji}  S. Iijima, Nature {\bf 354}, 56 (1991); A. Thess{\it \ et al}%
., Science {\bf 273}, 483 (1996).

\bibitem{egger}  R. Egger {\it et al.}, cond-mat/0008008.

\bibitem{morp}  A. F. Morpurgo {\it et al.}, Science {\bf 286}, 263 (1999).

\bibitem{tomo}  S. Tomonaga, Prog. Theor. Phys. (kyoto) {\bf 5}, 544 (1950).

\bibitem{lutt}  J. M. Luttinger, J. Math. Phys. {\bf 4}, 1154 (1963).

\bibitem{hald}  F. D. M. Haldane, PRL {\bf 47}, 1840 (1981). {\bf \ }

\bibitem{voit}  J. Voit, Rep. Prog. Phys. {\bf 57}, 977 (1994);
cond-mat/0005114.

\bibitem{rao}  S. Rao and D. Sen, cond-mat/0005492.

\bibitem{furu}  A. Furusaki, Phys. Rev. {\bf B\ 57}, 7141 (1998-II).

\bibitem{kane}  C. L. Kane and M. P. A. Fisher, Phys. Rev. {\bf B 46}, 15
233 (1992).

\bibitem{furu2}  A. Furusaki and N. Nagaosa, Phys. Rev. {\bf B 47}, 3827
(1993).

\bibitem{cham}  C. de C. Chamon and X. G. Wen, Phys. Rev. Lett. {\bf 70},
2605 (1993).

\bibitem{haug}  H. Haug and A. -P. Jauho, {\sl Quantum Kinetics in Transport
and Optics of Semiconductors} (Springer-Verlag, Berlin, 1998).

\bibitem{meir}  Y. Meir et al., Phys. Rev. Lett. {\bf 66}, 3048 (1991); L.
Wang {\it et al.}, Phys. Rev. Lett., {\bf 73}, 585 (1994).

\bibitem{grad}  I. S. Gradshteyn and I. M. Ryzhik, {\sl Table of Integrals,
Series, and Products} (Academic Press, Inc., New York, 1980).

\bibitem{expl}  These conditions are different from that of Chamon and Wen,%
\cite{cham} where the current formula stands down to zero temperature for
interacting parameter $g<\frac 12$. In our calculation, the imaginary part
of the self-energy has been neglected, which will play important role in the
tunneling process when $T\rightarrow 0$.

\bibitem{ausla}  O. M. Auslaender {\it et al.}, Phys. Rev. Lett. {\bf 84},
1764 (2000).
\end{references}
\end{document}